\documentclass[twocolumn,showpacs,english,pra,amssymb]{revtex4}
\usepackage[T1]{fontenc}
\usepackage[latin1]{inputenc}
\usepackage{babel}
\usepackage{graphicx}

\makeatletter

\providecommand{\LyX}{L\kern-.1667em\lower.25em\hbox{Y}\kern-.125emX\@}
\let\SF@@footnote\footnote
\def\footnote{\ifx\protect\@typeset@protect
    \expandafter\SF@@footnote
  \else
    \expandafter\SF@gobble@opt
  \fi
}
\expandafter\def\csname SF@gobble@opt \endcsname{\@ifnextchar[
  \SF@gobble@twobracket
  \@gobble
}
\edef\SF@gobble@opt{\noexpand\protect
  \expandafter\noexpand\csname SF@gobble@opt \endcsname}
\def\SF@gobble@twobracket[#1]#2{}

\usepackage{graphics}
\usepackage{amsmath}
\usepackage{epsfig}
\usepackage{bm}

\makeatother
\begin{document}

\title{Gauge P representations for quantum-dynamical problems: \\
 Removal of boundary terms}

\author{P. Deuar}

\email{deuar@physics.uq.edu.au}

\author{P. D. Drummond}

\email{drummond@physics.uq.edu.au}

\homepage{www.physics.uq.edu.au/BEC}

\affiliation{Department of Physics, University of Queensland, QLD 4072, Brisbane,
Australia}

\date{\today{}}

\begin{abstract}
P representation techniques, which have been very successful in quantum
optics and in other fields, are also useful for general bosonic quantum
dynamical many-body calculations such as Bose-Einstein condensation. We introduce a representation
called the gauge P representation which greatly widens the range of
tractable problems. Our treatment results in an infinite set of possible
time-evolution equations, depending on arbitrary gauge functions that
can be optimized for a given quantum system. In some cases, previous
methods can give erroneous results, due to the usual assumption of
vanishing boundary conditions being invalid for those particular systems.
Solutions are given to this boundary-term problem for all the cases
where it is known to occur: two-photon absorption and the single-mode
laser. We also provide some brief guidelines on how to apply the stochastic
gauge method to other systems in general, quantify the freedom of
choice in the resulting equations, and make a comparison to related
recent developments. 
\end{abstract}


\maketitle

\section{Introduction}

\label{INTRO} One of the most difficult problems in theoretical physics
is also conceptually the simplest. How does one calculate the dynamical
time evolution or even the ground state of an interacting many-body
quantum system? In essence, this is a natural part of practically
any comparison of quantum theory with experiment. The difficulty is
that the Hilbert space of all but the most trivial cases can be enormous.
This implies that a finite computer is needed to to solve problems
that can easily become nearly infinite in dimensionality, if treated
using an orthogonal basis expansion.

In this paper, we formally introduce and give examples of techniques
for treating general bosonic many-body quantum systems, which we call
gauge P representations. These are an extension of the phase-space
method called the positive-P representation \cite{DG-PosP}, and have
been recently used in the context of interacting Bose gases \cite{ccp2k,canonical}.
The advantages of the new technique are the following.

(1) The elimination of certain types of mathematical terms known as boundary-term
corrections, which have caused problems in the positive-P representation
for over a decade \cite{SG-abs,SS-fail,GGD-Validity}. This is the
main focus of the present paper. 

(2) Greatly reduced sampling error in computations. Gauge P representations
have been used recently to reduce the sampling error in Kerr oscillator
simulations \cite{ccp2k}. 

(3) The extension of allowable problems to 'imaginary-time' canonical
ensemble calculations. These problems will be treated elsewhere. 

Related extensions to the positive-P representation --- although restricted
to the scalar interacting Bose gas problem --- have also been introduced
recently. Different procedures have been introduced by Carusotto,
Castin, and Dalibard \cite{Paris1,Paris2}, and by Plimak, Olsen, and
Collett \cite{Plimak}. These methods implicitly assume the absence
of boundary term corrections. This paper unifies and substantially
generalizes all these recent advances. It also shows how the gauge
method can be used to solve the long-standing problem of boundary-term
corrections in the positive P representation. Comparisons to the other
methods are given in an Appendix.

Owing to the work of Wilson \cite{wilson:74}, and many others \cite{Ceperley},
we know that large Hilbert space problems can often be treated using
stochastic or Monte Carlo techniques for the ground-state, particle
masses, and finite-temperature correlations. This is the basis for
much work in computational quantum statistical mechanics, and in QCD
as well. However, Wilson's and other related methods are restricted
to static or `imaginary-time' calculations, rather than quantum-dynamical
problems.

Methods like these that use orthogonal basis sets have not proven
useful for quantum dynamics; owing to the notorious phase problem
that occurs when trying to sum over families of paths in real-time
Feynman path integrals. For this reason, the many-body quantum time-evolution
problem is often regarded as inherently insoluble due to its exponential
complexity. In fact, it was this very problem that motivated the original
proposal of Feynman \cite{feynman:82} to develop quantum computers.
In these (usually conceptual) devices, the mathematical problem is
solved by  a physical system consisting
of evolving `qubits' or two-state physical devices. Fortunately, this
method of doing calculations is not the only one, since no large enough
quantum computer exists at present\cite{Factor15}.

Historically, an alternative route is the use of quasi-probability
representations of the quantum state, which either implicitly or explicitly
make use of a non orthogonal basis. The term quasi-probability is
used because there can be no \emph{exact} mapping of all quantum states
to a classical phase space with a positive distribution \cite{Neumann}
that also preserves all the marginal probabilities. These methods
include the Wigner \cite{Wig-Wigner} (W), Glauber-Sudarshan (P) \cite{Gla-P,Sud-P},
and Husimi (Q) \cite{Hus-Q,CG-Q} representations. The classical phase-space
representations can be classified according to the operator ordering
that stochastic moments correspond to: the W is symmetrically ordered,
the Q is anti-normally ordered, while the P representation is normally
ordered. Apart from numerous laser physics and quantum optics calculations,
these methods have also been used to some extent in quantum statistical
mechanics: for example, the theory of BEC phase fluctuations \cite{becpf}.

None of these methods result in a stochastic time evolution with a
positive propagator when there are nonlinearities. To achieve this,
a better approach is to use a non-classical phase space of higher
dimension. A complex higher-dimensional `R representation' was proposed
in Glauber's seminal paper on coherent state expansions \cite{Gla-P}.
The first probabilistic method of this type was the positive-P representation \cite{DG-PosP}
(+P), which has proved capable of performing stochastic time-domain
quantum calculations in some many-body quantum systems \cite{carterdrummond:87}.
This uses a basis of coherent states that are not orthogonal, thus
allowing freedom of choice in the construction of the representation.
The positive-P representation of a quantum state is therefore the
most versatile out of a large group of quasi-probability distributions
developed to aid quantum mechanical calculations. It has been successfully
applied to mesoscopic systems such as quantum solitons \cite{carterdrummond:87,qsol2,qsol3}
and the theory of evaporative cooling \cite{evapcool}, which correctly
reproduces the formation of a BEC --- as observed in experiment \cite{obsex1,obsex2,obsex3}.

Quasi-probability distributions of this type are computationally superior
to direct density matrix methods, which are susceptible to computational
complexity blow-up for large Hilbert spaces. Provided certain boundary
terms vanish, the usual procedure is to generate a Fokker-Planck equation
(which will vary depending on the distribution chosen) from the master
equation, and then to convert this to a set of stochastic Langevin
equations. For some simple cases, it may even be possible to arrive
at appealing results directly from the Fokker-Planck equation (FPE). The
resulting stochastic equations can be thought of just as quantum mechanics
written in different variables. They have two main advantages over
orthogonal basis-state methods, as follows.

First, the whole quantum dynamics can be written exactly in terms
of a small number of stochastic equations. In a one-mode case, there
is just one complex variable for P and Q and W, and two complex variables
for +P. Although a simulation requires us to average over many realizations
of the stochastic process, this is often more practical than solving
the infinite set of deterministic equations required to solve directly
for all the elements of a density matrix. Such an infinite set may
be truncated, but this is only a good approximation for a system with
few particles, and no more than a few modes. 

Second, for a many-mode problem the Hilbert space dimension is
\( N=n^{M} \) for the case of \( n \) particles distributed over
\( M \) modes. This gives exponential growth as a function of the
number of modes. However, the number of quasi-probability dynamical
equations grows only \textit{linearly} with the number of modes, rather
than \textit{exponentially} in the case of direct methods. Other stochastic
methods, known as quantum-trajectory methods, can be used to reduce
the \( N^{2} \) dimensionality of an \( N\times N \) density matrix
problem to that of the \(N\)-dimensional underlying Hilbert space
--- but this is clearly insufficient to solve the complexity problem
inherent in the exponential growth of the Hilbert space dimension. 

There are, however, some caveats when using these distributions. In
particular, the vanishing of boundary terms is an important fundamental
issue with quasi-probability distributions, and it is this issue that
we focus on mostly in this paper. To get an overall picture, consider
that once we have a time-evolution problem there are five typical
requirements that are encountered in deriving stochastic equations
for quasi-probability representations of many-body systems. These
requirements occur in closed (unitary evolution) systems, in open
systems (in general described by a master equation), or even using
a distribution to solve for the canonical ensemble in imaginary time.
As such, these requirements are generic to the use of stochastic equations
with operator representations:

(1) \textit{Positive distribution}. A well-behaved positive distributions
for all quantum states, including especially the chosen initial condition,
is essential for a general algorithm. For example, a number state
has a highly singular P distribution, and a W distribution that is
negative in some regions of phase space \cite{KS-70}, making either
distribution impossible to interpret probabilistically for these states.
The R distribution is inherently complex. Such problems do not occur
for the Q or +P representation --- these are positive, and well-behaved
for all quantum states \cite{DG-PosP}. 

(2) \emph{Ultraviolet convergence}. While normally-ordered representations
are well behaved at large momentum, non-normally-ordered representations
of quantum fields --- such as the Q or W representations --- typically face
the problem of ultraviolet divergence in the limit of large momentum
cutoff \cite{evapcool}. This means that almost any observable quantity
will involve the simulation of a (nearly) infinitely noisy classical
field, leading to diverging standard deviations in two or more space
dimensions, even for linear systems. This rules out the Q and W distributions
for quantum field simulations in higher than one-dimensional environments. 

(3) \textit{Second-order derivatives}. Only FPEs with second or infinite-order
derivatives can be translated into stochastic equations \cite{Arnold}.
Normally-ordered methods such as the P and +P representations can handle
most commonly occurring nonlinearities and two-body interactions,
with only second-order derivatives. Non-normally ordered representations
of quantum fields often lead to third- or higher-order partial derivatives
in the Fokker-Planck equation with no stochastic equivalent. For example,
the Wigner representation gives such problems for almost any nonlinear
term in the master equation. 

(4) \textit{Positive-definite diffusion}. A Fokker-Planck equation must
have positive-definite diffusion, to allow simulation with stochastic
processes \cite{Arnold}. When the master equation has nonlinear terms,
this does not occur with any of the classical representations. However,
the +P representation is guaranteed to always produce positive-definite
diffusion \cite{DG-PosP}, provided no higher derivative terms occur. 

(5) \textit{Vanishing boundary terms}. In the derivation of the Fokker-Planck
equations, it is assumed that certain boundary terms arising in partial
integration can be neglected. This is not always the case. Boundary
terms due to power-law tails can occur when there are moving singularities
that can escape to infinity in finite time. In the +P method, such
trajectories may cause systematic errors in stochastic averages \cite{GGD-Validity},
especially for non-integrable dynamical systems. These problems are
exponentially suppressed when linear damping rates are increased,
but can be large at low damping. 

The +P method is often the representation of choice, because it satisfies
conditions one to four. Gauge representations (G) combined with stochastic
methods to be treated in this paper, share these advantages with the
+P representation. However, they can also satisfy the fifth requirement
--- for an appropriate gauge choice --- hence allowing all of the mathematical
problems in simulating time evolution to be treated. For this reason,
the present paper will focus on solving boundary-term issues encountered
with the +P representation for certain nonlinear master equations.
The overall picture is summarized in Table~\ref{THETABLE}, as applied
to the two-boson anonlinear absorber cases
treated here in Sec.~\ref{D12PD}:

We emphasize that the particular examples treated here have a small
particle number and extremely low (or zero) linear damping. As such,
they are soluble using other techniques, which allows us to test the
accuracy of gauge techniques. Our purpose is to demonstrate the success
of the stochastic gauge method in simple cases where boundary terms
arise within the +P representation. In this way, we can understand
more complex situations where no exact result is known.

We will first derive and describe the stochastic gauge method in
Secs.~\ref{GAUGES} and ~\ref{GF},
and subsequently work through two examples: First, solving the boundary-value 
problem for the driven one- and two-photon absorber in Sec.~\ref{D12PD}.
Second, in Sec.~\ref{LAS} we will consider the one-mode laser
at extremely low power, which exhibits boundary term errors when very
non-optimal starting conditions are used. This example will show that
gauge methods can also be used to remove errors from this system,
but some judgment must be employed to avoid choosing a pathological
initial distribution. In the Appendix, we compare the methods derived
here with recent related extensions of the positive-P representation
by Carusotto and co-workers \cite{Paris1,Paris2}, Plimak \textit{et
al.} \cite{Plimak}, and Deuar and Drummond \cite{ccp2k}. 

\begin{table*}
\caption{Comparison of phase-space representations as applied to stochastic treatments
of a one- and two- boson nonlinear absorber.\label{THETABLE}}
\begin{ruledtabular}
\begin{tabular}{cccccccc}
Method&
 Form of&
 UV&
 Order of&
 Non-negative&
 Stochastic&
 Boundary term&
 Simulated\\
&
Distribution&
 converges&
 derivatives&
 diffusion&
 simulations&
 removal&
 correctly\\
\hline
W&
 Real&
 No&
 4&
 Sometimes&
 No&
  &
  \\
Q&
 Positive&
 No&
 4&
 Yes&
 No&
  &
  \\
R&
 Complex&
 Yes&
 2&
  &
 No&
  &
  \\
P&
 Singular&
 Yes&
 2&
 No&
 No&
  &
  \\
+P&
 Positive&
 Yes&
 2&
 Yes&
 Yes&
 No&
 Sometimes\\
G&
 Positive&
 Yes&
 2&
 Yes&
 Yes&
 Yes&
 Yes  \\
\end{tabular}
\end{ruledtabular}
\end{table*}

Finally, we point out a sixth requirement of containing the \textit{growth
of sampling error}: the averages calculated from the stochastic Langevin
equations correspond to quantum mechanical expectation values only
in the limit of infinitely many trajectories. Provided boundary terms
do not occur, the averages will approach the correct values --- within
an acceptable sampling error --- for sufficiently many trajectories.
If this number should increase rapidly with time, the simulation will
only be of use for a limited period  \cite{ccp2k}. 

The problem of growing sampling error can occur even when there are
no boundary terms, and may be regarded as the ultimate frontier in
representation theory, just as similar issues dominate the theory
of classical chaos. This is less of a fundamental issue, since the
sampling error can always be estimated and controlled by increasing
the number of trajectories. This is simply a matter of moving to a
clustered, parallel computational model, or repeating the calculation
many times. Nevertheless, it is of great practical significance. The
sampling error problem requires careful gauge optimization, and remains
an open area for investigation. An intelligent choice of gauge can
often vastly outweigh a brute force computational approach, in terms
of sampling error.

\section{Gauge Operator Representations}

\label{GAUGES}

In gauge representations, the density matrix to be computed is expanded
in terms of a coherent state basis. For definiteness, we shall focus
on the coherent states of the harmonic oscillator, which are useful
in expanding Bose fields; but other choices are clearly possible.
The expansion kernel is more general than that used in the positive-P 
representation. In order to define the notation, we start by introducing
a set of boson annihilation and creation operators \( \widehat{a}_{i} \)
, \( \widehat{a}_{i}^{\dagger } \). The operator \( \widehat{n}_{i}=\widehat{a}_{i}^{\dagger }\widehat{a}_{i} \)
is therefore the boson number operator for the \(i\)th mode or
site. Boson commutation relations of \( [\widehat{a}_{i},\widehat{a}_{j}^{\dagger }]=\delta _{ij} \)
hold for the annihilation and creation operators.

\subsection{Coherent states}

If \( \bm {\alpha }=(\alpha _{1},\dots ,\alpha _{M}) \) is a complex
\( M \)-dimensional vector with \( \alpha _{i}=x_{i}+iy_{i} \),
and \( \widehat{\mathbf{a}}=(\widehat{a}_{1},\dots ,\widehat{a}_{M}) \)
is an \( M \)-dimensional vector of annihilation operators,
then the Bargmann coherent state \( \left\Vert \bm {\alpha }\right\rangle  \)
is defined by \begin{equation}
\left\Vert \bm {\alpha }\right\rangle =\exp \left[ \bm {\alpha }\cdot \widehat{\mathbf{a}}^{\dagger }\right] \left| 0\right\rangle =\exp \left[ |\bm {\alpha }|^{2}/2\right] \left| \bm {\alpha }\right\rangle \, ,
\end{equation}

where \( \left| \bm {\alpha }\right\rangle  \) is the usual normalized
coherent state which is a simultaneous eigenstate of all the annihilation
operators. The inner product of two Bargmann coherent states is 
\begin{equation}
\left\langle \bm {\beta }^{*}\right. \left\Vert \bm {\alpha }\right\rangle =\exp \left[ \bm {\alpha }\cdot \bm {\beta }\right] \, \, .
\end{equation}

It is important to notice here that \( \left\Vert \bm {\alpha }\right\rangle  \)
is an analytic function of the complex vector \( \bm {\alpha } \)
. The following identities therefore follow immediately:

\begin{eqnarray}
\widehat{a}_{i}\left\Vert \bm {\alpha }\right\rangle  & = & \alpha _{i}\left\Vert \bm {\alpha }\right\rangle \nonumber \\
\widehat{a}_{i}^{\dagger }\left\Vert \bm {\alpha }\right\rangle  & = & \frac{\partial }{\partial \alpha _{i}}\left\Vert \bm {\alpha }\right\rangle \, \, .
\end{eqnarray}

Since \( \left\Vert \bm {\alpha }\right\rangle  \) is an analytic function,
the notation \( \partial /\partial \alpha _{i} \) is interpreted
here as an analytic derivative, which can be evaluated in either the
real or imaginary directions,\begin{equation}
\frac{\partial }{\partial \alpha _{i}}\left\Vert \bm {\alpha }\right\rangle =\frac{\partial }{\partial x_{i}}\left\Vert \bm {\alpha }\right\rangle =-i\frac{\partial }{\partial y_{i}}\left\Vert \bm {\alpha }\right\rangle \, \, .
\end{equation}

Since the coherent states are an over-complete basis set, any operator
can be expanded in more than one way using coherent states. For example,
the simplest resolution of the identity operator is

\begin{equation}
\widehat{I}=\frac{1}{\pi ^{M}}\int \left| \bm {\alpha }\right\rangle \left\langle \bm {\alpha }\right| d^{2M}\bm {\alpha }.
\end{equation}

Thus, introducing a second \( M \)-dimensional vector \( \bm {\beta } \),
we can expand any operator \( \widehat{O} \) directly as\begin{eqnarray}
\widehat{O} & = & \frac{1}{\pi ^{2M}}\int \int \left| \bm {\alpha }\right\rangle \left\langle \bm {\alpha }\right| \widehat{O}\left| \bm {\beta ^{*}}\right\rangle \left\langle \bm {\beta ^{*}}\right| d^{2M}\bm {\alpha }d^{2M}\bm {\beta }\, \nonumber \\
 & = & \int \int O(\bm {\alpha },\bm {\beta })\left| \bm {\alpha }\right\rangle \left\langle \bm {\beta ^{*}}\right| d^{2M}\bm {\alpha }d^{2M}\bm {\beta }\, \, .
\end{eqnarray}

Here, we have introduced\begin{equation}
\label{O-canonical}
O(\bm {\alpha },\bm {\beta })=\frac{1}{\pi ^{2M}}\left\langle \bm {\alpha }\right| \widehat{O}\left| \bm {\beta ^{*}}\right\rangle \, \, .
\end{equation}

\subsection{P representations}

The possibility of expanding any operator in terms of coherent states
leads to the idea that such an expansion can be used to calculate
observable properties of a quantum density matrix \( \widehat{\rho } \)
. Historically, this was first proposed by Glauber and Sudarshan \cite{Gla-P,Sud-P},
who suggested a diagonal expansion of the form\begin{equation}
\widehat{\rho }=\int P(\bm {\alpha })\left| \bm {\alpha }\right\rangle
\left\langle \bm {\alpha }\right| d^{2M}\bm {\alpha }\  .
\end{equation}
 Unlike the direct expansion given above, this has no off-diagonal
elements. Surprisingly, expansions of this type always exist, as long
as the function \( P(\bm {\alpha }) \) is defined to allow highly
singular generalized functions and non-positive distributions \cite{KS-70}.

As these do not have a stochastic interpretation, the positive-P representation
was introduced \cite{DG-PosP}, which is defined as\begin{equation}
\widehat{\rho }=\int P^{(+)}(\bm {\alpha },\bm {\beta })\frac{\left| \bm {\alpha }\right\rangle \left\langle \bm {\beta ^{*}}\right| }{\left\langle \bm {\beta ^{*}}\right. \left| \bm {\alpha }\right\rangle }\, d^{2M}\, \bm {\alpha }d^{2M}\bm {\beta }\, 
\end{equation}
 for an \( M \)-mode system.

It is always possible to obtain an explicitly positive-definite distribution
of this type \cite{DG-PosP}, with the definition \begin{eqnarray}
P^{(+)}(\bm {\alpha },\bm {\beta }) & = & \frac{1}{(4\pi ^{2})^{M}}\exp \left[ -\left| \frac{\bm {\alpha }-\bm {\beta ^{*}}}{2}\right| ^{2}\right] \nonumber \\
 &  & \quad \times \quad \left\langle \frac{\bm {\alpha }+\bm {\beta
^{*}}}{2}\right| \widehat{\rho }\left| \frac{\bm {\alpha }+\bm {\beta
^{*}}}{2}\right\rangle \, \, .\ \ \label{CanonicalP} 
\end{eqnarray}
 This form always exists, as do an infinite class of equivalent positive
distributions. Even simpler ways to construct the positive-P representation
are available in some cases. For example, if the Glauber-Sudarshan
representation exists and is positive, then one can simply construct\begin{equation}
P^{(+)}(\bm {\alpha },\bm {\beta })=P(\bm {\alpha })\delta ^{2M}(\bm {\alpha }-\bm {\beta ^{*}})\, \, .
\end{equation}
 The stochastic time evolution of the positive-P distribution does
not generally preserve the above compact forms, and may allow less
compact positive solutions instead. However, to obtain a time-evolution
equation, it is necessary to use partial integration, with the assumption
that boundary terms at infinity can be neglected. It is these less
compact solutions, occurring during time evolution with a nonlinear
Fokker-Planck equation, that lead to power-law tails in the distribution
--- and hence boundary-term problems caused by the violation of the
assumption that these terms vanish.

\subsection{Gauge representations}

A technique for constructing an even more general positive distribution
is to introduce a quantum complex amplitude \( \Omega  \), which
can be used to absorb the quantum phase factor. This leads to the
result that any Hermitian density matrix \( \widehat{\rho } \) can
be expanded in an over-complete basis \( \widehat{\Lambda }(\overrightarrow{\alpha }) \),
where \( \overrightarrow{\alpha }=(\Omega ,\, \bm {\alpha },\bm {\beta }) \),
and \begin{eqnarray}
\widehat{\Lambda }(\overrightarrow{\alpha }) & = & \Omega \frac{\left\Vert \bm {\alpha }\right\rangle \left\langle \bm {\beta }^{*}\right\Vert }{\left\langle \bm {\beta }^{*}\right. \! \left\Vert \bm {\alpha }\right\rangle }\nonumber \\
 & = & \Omega \left\Vert \bm {\alpha }\right\rangle \left\langle \bm {\beta }^{*}\right\Vert \exp \left[ -\bm {\alpha }\cdot \bm {\beta }\right] \, \, .\label{lambda-defn} 
\end{eqnarray}
 We define the gauge representation \( G(\overrightarrow{\alpha }) \)
as a real, positive function that satisfies the following equation:

\begin{eqnarray}
\widehat{\rho } & = & \int G(\overrightarrow{\alpha })\left[ \widehat{\Lambda }(\overrightarrow{\alpha })\right] d^{4M+2}\overrightarrow{\alpha }\, \nonumber \\
 & = & \frac{1}{2}\int G(\overrightarrow{\alpha })\left[ \widehat{\Lambda }(\overrightarrow{\alpha })+\text{H.c.}\right] d^{4M+2}\overrightarrow{\alpha }\, .
\end{eqnarray}
 The last line above follows from the fact that \( \widehat{\rho } \)
is a Hermitian density matrix and \( G(\overrightarrow{\alpha }) \)
is real. Here, \( \text{H.c.} \) is used as an abbreviation for Hermitian
conjugate. The use of a complex weight in the above gauge representation
is similar to related methods introduced recently for interacting
Bose gases \cite{Paris1,Paris2}, except that we multiply the weight
by a normalized (positive-P) projector, in order to simplify the resulting
algebra.

As an existence theorem that shows that this representation always
exists, consider the complex solution \begin{equation}
P_{0}(\bm {\alpha },\bm {\beta })=\frac{1}{\pi ^{2M}}\left\langle \bm {\alpha }\right| \widehat{\rho }\left| \bm {\beta ^{*}}\right\rangle \left\langle \bm {\beta ^{*}}\right. \left| \bm {\alpha }\right\rangle \, \, 
\end{equation}
obtained from Eq. (\ref{O-canonical}), with a phase \( \theta =\arg (P_{0}) \)
, and simply define\begin{equation}
G(\overrightarrow{\alpha })=\left| P_{0}(\bm {\alpha },\bm {\beta })\right| \delta ^{2}(\Omega -\exp [i\theta (\bm {\alpha },\bm {\beta })])\, \, .
\end{equation}

In this type of gauge representation, \( G(\overrightarrow{\alpha }) \)
is a positive distribution over a set of Hermitian density-matrix
elements \( \widehat{\Lambda }+\widehat{\Lambda }^{\dagger } \).
It is simple to verify that, by construction \begin{equation}
\text{Tr}\left( \widehat{\Lambda }\right) =\Omega \, \, .
\end{equation}

For the case of \( \Omega =1 \), this representation reduces to the
positive-P representation, and the kernel \( \widehat{\Lambda }(\overrightarrow{\alpha }) \)
is a projection operator. Since the positive-P representation is a
complete representation, it follows that another way to construct
the gauge P representation is always available, if one simply
defines \begin{equation}
G(\overrightarrow{\alpha })=P^{(+)}(\bm {\alpha },\bm {\beta })\delta ^{2}(\Omega -1)\, \, .
\end{equation}

As a simple example, a thermal ensemble with \( n_{0} \) bosons per
mode gives a diagonal P distribution that is Gaussian, so that \begin{equation}
G_{th}(\overrightarrow{\alpha })\propto \exp \left[ -\left| \bm {\alpha }\right| ^{2}/n_{0}\right] \delta ^{2M}(\bm {\alpha }-\bm {\beta ^{*}})\delta ^{2}(\Omega -1)\, \, .
\end{equation}

One advantage of the proposed representation is that it allows more general
expansions than the positive-P distribution, and also includes the
case of the complex P representation --- which has proved useful
in solving for non-equilibrium steady-states in quantum systems.

\subsection{Operator identities}

The utility of these methods arises when they are used to calculate
time (or imaginary time --- for which the positive-P distribution cannot
be used) evolution of the density matrix. This occurs via a Liouville
equation of generic form

\begin{equation}
\frac{\partial }{\partial t}\widehat{\rho }=\widehat{L}(\widehat{\rho })\, \, ,
\end{equation}
 where the Liouville superoperator typically involves pre- and post-multiplication
of \( \widehat{\rho } \) by annihilation and creation operators.
As an example, the equation for purely unitary time evolution under
a Hamiltonian \( \widehat{H} \) is \begin{equation}
i\hbar \frac{\partial }{\partial t}\widehat{\rho }=\left[ \widehat{H},\widehat{\rho }\right] \, \, .
\end{equation}

Effects of the annihilation and creation operators on the projectors
are obtained using the results for the actions of operators on the
Bargmann states, \begin{eqnarray}
\widehat{\mathbf{a}}\widehat{\Lambda }(\overrightarrow{\alpha }) & = & \bm {\alpha }\widehat{\Lambda }(\overrightarrow{\alpha })\nonumber \\
\widehat{\mathbf{a}}^{\dagger }\widehat{\Lambda }(\overrightarrow{\alpha }) & = & \left[ \bm {\partial }_{\alpha }+\bm {\beta }\right] \widehat{\Lambda }(\overrightarrow{\alpha })\nonumber \\
\widehat{\Lambda }(\overrightarrow{\alpha }) & = & \Omega \, \partial _{\Omega }\, \widehat{\Lambda }(\overrightarrow{\alpha })\, \, .
\end{eqnarray}
 For brevity, we use \( \overrightarrow{\partial }=(\partial _{\Omega },\bm {\partial }_{\alpha },\bm {\partial }_{\beta }) \)
to symbolize either \( \left( \partial ^{x}_{i}\equiv \partial /\partial x_{i}\right)  \)
or \( -i\left( \partial ^{y}_{i}\equiv \partial /\partial y_{i}\right)  \)
for each of the \( i=0,\dots, 2M \) complex variables \(
\overrightarrow{\alpha } \). 
This is possible since \( \widehat{\Lambda }(\overrightarrow{\alpha }) \)
is an analytic function of \( \overrightarrow{\alpha } \), and an
explicit choice of the derivative will be made later.

Using the operator identities given above, the operator equations
can be transformed to an integro-differential equation,\begin{eqnarray}
\frac{\partial \widehat{\rho }}{\partial t} & = & \int G(\overrightarrow{\alpha })\left[ \mathcal{L}_{A}\widehat{\Lambda }(\overrightarrow{\alpha })\right] d^{4M+2}\overrightarrow{\alpha }\, \, .
\end{eqnarray}
 Here the anti-normal ordered notation \( \mathcal{L}_{A} \) indicates
an ordering of all the derivative operators to the right. As an example,
in the Hamiltonian case, if the original Hamiltonian \( \widehat{H}(\widehat{\mathbf{a}},\widehat{\mathbf{a}}^{\dagger }) \)
is normally-ordered (annihilation operators to the right), then \begin{equation}
\label{Differential}
\mathcal{L}_{A}=\frac{1}{i\hbar }\left[ H_{A}(\bm {\alpha },\bm {\partial }_{\alpha }+\bm {\beta })-H_{A}(\bm {\beta },\bm {\partial }_{\beta }+\bm {\alpha })\right] \, \, .
\end{equation}
 If no terms higher than second order occur, this procedure gives
a differential operator with the following general expansion:

\begin{equation}
\mathcal{L}^{(+)}_{A}=V+A^{(+)}_{j}\partial _{j}+\frac{1}{2}D_{ij}\partial _{i}\partial _{j}\, .
\end{equation}
 where, to simplify notation, the Latin indices \( i,j,k \) will
from now on be summed over \( i=1,\dots,2M \), since no derivatives with
respect to \( \Omega  \) are used as yet. V is a term not involving derivative operators
 with respect to any of the variables in \(\overrightarrow{\alpha}.\) The drift term \( A^{(+)}_{j} \)
that is normally found using the positive-P representation is labeled
with the superscript \( (+) \) to identify it.

At this stage, the usual procedure in representation theory is to
integrate by parts, provided boundary terms vanish. This gives a normally-ordered
differential operator acting on the distribution itself, of form

\begin{equation}
\frac{\partial }{\partial t}G(\overrightarrow{\alpha })=\left[ V-\partial _{j}A^{(+)}_{j}+\frac{1}{2}\partial _{i}\partial _{j}D_{ij}\right] G(\overrightarrow{\alpha })\, .
\end{equation}

This type of generalized Fokker-Planck equation can be treated formally
using techniques developed by Graham, involving time-symmetric curved-space
path integrals \cite{Path-int}. For computational purposes, we require
special choices of the analytic derivatives to obtain a positive-definite
diffusion, so that the path integrals have equivalent stochastic equations \cite{Arnold}.
We emphasize here that the equations resulting are quite different
to those obtained from the direct insertion of a coherent state identity
into a Feynman path integral --- which results in severe convergence
problems \cite{Phase-path}. The usual positive-P representation equations
are obtained at this stage --- provided there is no potential term ---
and can be transformed to stochastic equations using the techniques
described in the following section.

\section{Gauge functions}
\label{GF}
In gauge representations, the time evolution of the representation
is modified from the usual positive-P representation equations, by
the introduction of a number of arbitrary and freely defined functions
on the phase space. This freedom of choice is, of course, not present
with an orthogonal basis, and is due to the non-orthogonal nature
of a coherent basis set. Although we do not investigate other cases,
it is worth noting that a similar gauge freedom is implicitly present
whenever a non-orthogonal expansion is used --- even if it involves
different states from the choice of coherent states made here (e.g.,
the Fock state wave functions in Refs. \cite{Paris1,Paris2}).

\subsection{Diffusion gauges}

We first introduce the diffusion gauges, which were implicitly present
in the original positive-P representation, but were only recognized
recently as allowing improvements in the sampling error. These gauges
occur via the non-unique decomposition of the complex diffusion matrix
\( D \), which determines the stochastic correlations in the final
equations. Arbitrary functional parameters can therefore be inserted
into the final stochastic equations in the noise coefficients, which
may lead to further optimization of the simulation. This is because
the decomposition of the complex diffusion matrix \( D=BB^{T} \)
, which is needed to define a stochastic process, does not specify
the resulting noise matrix \( B \) completely.

It has been recently shown by Plimak, Olsen and Collett \cite{Plimak}
that for the Kerr oscillator using a decomposition different from
the obvious diagonal one leads to impressive improvements in the signal-to-noise 
ratio of the simulation (briefly described in Appendix~\ref{PLIMAKA}).
This somewhat surprising result leads us to try to quantify the amount
of freedom of choice available from this source.

Since \( D=D^{T} \), it can always be diagonalized by a complex orthogonal
transformation \begin{equation}
\label{Bmat}
D=O\lambda ^{2}O^{T}=B^{(+)}B^{(+)T}\, \, ,
\end{equation}
 where \( \lambda  \) is the diagonal matrix whose square gives the
eigenvalues of \( D \). Thus \( B^{(+)}=O\lambda  \) can be considered
the canonical, or {}``obvious'' choice of decomposition, unique
apart from the \( 2M \) signs of the diagonal terms. However, for
any orthogonal \( U \), if \( B^{(+)} \) is a valid decomposition
of \( D \), then so is the matrix \( B=B^{(+)}U \). Hence, any matrix
in the whole orthogonal family \( B=O\lambda U \) is a valid decomposition.
This can be easily quantified using a basis \[
\sigma ^{(ij)}_{kl}=\delta _{ik}\delta _{jl}-\delta _{il}\delta _{jk}\, \, ,\]
 of the \( M(2M-1) \) independent antisymmetric \( 2M\times 2M \)
matrices \( \underline{\underline{\sigma }}^{(ij)} \)~. One simply
introduces \begin{equation}
U=\exp \left( \sum _{i<j}g_{ij}(\overrightarrow{\alpha },t)\, \underline{\underline{\sigma }}^{(ij)}\right) \, \, .
\end{equation}
 As an example, for a one-mode case there is one complex gauge function
introduced this way, which is \( g^{d}=g_{12} \). The resulting
transformation is \begin{eqnarray}
U & = & \exp \left( g^{d}\, \underline{\underline{\sigma }}^{(12)}\right) \, \, \nonumber \\
 & = & \cos (g^{d})+\underline{\underline{\sigma }}^{(12)}\sin (g^{d})\, \, ,
\end{eqnarray}
where the antisymmetric matrix \( \underline{\underline{\sigma }}^{(12)} \)
is proportional to a Pauli matrix,\begin{equation}
\underline{\underline{\sigma }}^{(12)}=\left[ \begin{array}{ccc}
0 &  & 1\\
-1 &  & 0
\end{array}\right] \, \, .
\end{equation}
Hence, if the noise was diagonal in the canonical form, the transformed
(but equivalent) noise matrix becomes \begin{equation}
\label{transform-B}
B=\left[ \begin{array}{ccc}
\lambda _{11}\cos (g^{d}) &  & \lambda _{11}\sin (g^{d})\\
-\lambda _{22}\sin (g^{d}) &  & \lambda _{22}\cos (g^{d})
\end{array}\right] \, \, .
\end{equation}

Now, the \( 2M \)-dimensional (complex) orthogonal matrix family
contains \( M(2M-1) \) free complex parameters, so there are
\( M(2M-1) \) diffusion gauge functions \( g_{ij}(\overrightarrow{\alpha },t) \)
that one can choose arbitrarily. This represents a large class of
specific gauges that can be used directly in simulations, as opposed
to the conditions on noise correlations usually given elsewhere \cite{Plimak}.

As pointed out by Graham \cite{Path-int}, there is a close similarity
between the theory of curved-space metrics, and path integrals with
a space-varying diffusion matrix. In the present context, the space
is complex, and we have a family of gauges that are generated on taking
the matrix square root of the diffusion matrix. We have not yet used
this matrix square root, but this decomposition will be applied to
obtain positive-definite equations via the choice of analytic derivatives
made in the following sections.

The above holds for square noise matrices \(B_{s}\), but one is also free
to add more noise coefficients in the manner \( B_{Q} = \left[
B_{s},Q\right] \). Then 
\begin{equation}
B_{s}B_{s}^T = \tilde{D} = D - QQ^T\ ,
\end{equation}
and all the \(2MW\) coefficients in the \(2M \times W\) matrix \( Q\) are
additional arbitrary complex functions. The freedom in \(B_{s}\) is the
same as before [i.e. \(M(2M-1)\) independent complex gauge functions],
with the proviso that \(B_{s}\) is now given by \(O\tilde{\lambda}U\)
where the square of \(\tilde{\lambda}\) gives the eigenvalues of the modified matrix
\(\tilde{D}\). The matrix \(B_{s}\) would be unchanged if \(QQ^T\) were
set to zero,
although this choice of \(Q\) does not appear to be useful; it just
adds extra noise. In general it is not clear whether or not any advantage can
be gained by introducing the additional off-square gauge functions
contained in \(Q\).

If \( B \) is given a functional form dependent
on the phase-space variables, it may lead to additional terms in the
Stratonovich form of the equations, which are considered later in
this section. In this situation one must be careful not to introduce
additional boundary-term errors arising from an excessively rapid
growth of the noise gauges.

There is a subtlety here which one must take some care with. The complex
noise matrix \( B \) is not the matrix that usually appears in the
theory of stochastic equations. Instead, this matrix is subsequently
transformed into an `equivalent' stochastic form, by taking advantage
of the analyticity of the Bargmann states. This means that the effect
of the diffusion gauges on the final equations also makes use of the
non-uniqueness of the coherent basis set itself.

\subsection{Drift gauges}

While the diffusion gauges can control sampling error due to the correlations
of noise terms, they cannot eliminate boundary terms due to singular
trajectories in the drift equations. The extra variable \( \Omega  \)
allows the \( \partial _{\Omega } \) identity to be used to convert
any potential term \( V \) to a derivative term, and also to introduce
a stochastic gauge to stabilize the resulting drift equations. This
defines an infinite class of formally equivalent Fokker-Planck equations,
in a similar way to related procedures in QED and QCD. To demonstrate
this, we introduce \( 2M \) arbitrary complex drift gauge functions
\( \mathbf{g}=[\, g_{i}(\overrightarrow{\alpha },t)\, ] \), to give
a new differential operator \( \mathcal{L}_{GA} \) whose form differs
from the original \( \mathcal{L}^{(+)}_{A} \) by terms that vanish identically
when applied to the kernel \( \widehat{\Lambda }(\overrightarrow{\alpha }) \),\begin{equation}
\label{LAg}
\mathcal{L}_{GA}=\mathcal{L}^{(+)}_{A}+\left[ V+\frac{1}{2}\mathbf{g}\cdot \mathbf{g}\, \Omega \, \partial _{\Omega }+g_{k}B_{jk}\partial _{j}\right] \left[ \Omega \partial _{\Omega }-1\right] \, .
\end{equation}
 The total differential operator \( \mathcal{L}_{GA} \) has an anti-normal
Fokker-Planck form. Extending the drift and diffusion matrices to
include the extra variable \( \Omega  \), we can write this --- summing
repeated \( a,b,c \) indices over \( a=0,\dots,2M \) --- as \begin{equation}
\mathcal{L}_{GA}=\left[ A_{a}\partial _{a}+\frac{1}{2}D_{ab}\partial _{a}\partial _{b}\right] \, .
\end{equation}
 The \emph{total} complex drift vector is \( \overrightarrow{A}=(A_{0},A_{1},\dots,A_{2M}) \);
where \begin{eqnarray}
A_{0} & = & \Omega V\nonumber \\
A_{j} & = & A^{(+)}_{j}-g_{k}B_{jk}\, \, \, \, .
\end{eqnarray}

The new diffusion matrix \(
\underline{\underline{D}} \) with elements \( D_{ab} \) is not diagonal, but it can
be factorized. Explicitly, it is now a square \( (2M+1)\times (2M+1) \) complex matrix, given by

\begin{eqnarray}
\underline{\underline{D}} & = & \left[ \begin{array}{cc}
\Omega ^{2}\mathbf{gg}^{T} & \Omega \mathbf{g}B^{T}\\
B\mathbf{g}^{T}\Omega  & BB^{T}
\end{array}\right] \nonumber \label{gaugediffusion1} \\
 & = & \left[ \begin{array}{cc}
0 & \Omega \mathbf{g}\\
0 & B
\end{array}\right] \left[ \begin{array}{cc}
0 & 0\\
\Omega \mathbf{g}^{T} & B^{T}
\end{array}\right] =\underline{\underline{B}}\, \underline{\underline{B}}^{T}\, \, .
\end{eqnarray}

Thus, we now have a new stochastic noise matrix with one added dimension,\begin{equation}
\label{gauge_{d}iffusion2}
\underline{\underline{B}}=\left[ \begin{array}{cc}
0 & \Omega \mathbf{g}\\
0 & B
\end{array}\right] \, \, .
\end{equation}

The operator (\ref{LAg}) was chosen to give this form for \( \underline{\underline{B}} \),
so that the only change in noise is for the \( \Omega  \) variable.

\subsection{Positive-definite diffusion}

It is always possible to transform these second-derivative terms into
a positive semi-definite diffusion operator on a real space, which
is a necessary requirement for a stochastic equation. When \( \underline{\underline{D}}=\underline{\underline{B}}\, \underline{\underline{B}}^{T} \),
divide \( \underline{\underline{B}}=\underline{\underline{B}}^{x}+i\underline{\underline{B}}^{y} \)
into its real and imaginary parts. A similar procedure is followed
for \( \overrightarrow{A} \).

Recalling that the original kernel was analytic, thus allowing for
more than one choice of derivatives, the choice for \( \partial _{a} \)
can now be made definite by choosing it so that the resulting drift and
diffusion terms are always real,\begin{eqnarray}
A_{a}\partial _{a} & \rightarrow  & A_{a}^{x}\partial ^{x}_{a}+A^{y}_{a}\partial ^{y}_{a}\, ,\\
D_{ab}\partial _{a}\partial _{b} & \rightarrow  & B_{ac}^{x}B_{bc}^{x}\partial ^{x}_{a}\partial ^{x}_{b}+B_{ac}^{y}B_{bc}^{x}\partial ^{y}_{a}\partial ^{x}_{b}+(x\leftrightarrow y)\, .\nonumber 
\end{eqnarray}
 Hence, the gauge differential operator can now be written explicitly
as \begin{equation}
\mathcal{L}_{GA}=\left[ \widetilde{A}_{\mu }\partial _{\mu }+\frac{1}{2}\widetilde{D}_{\mu \nu }\partial _{\mu }\partial _{\nu }\right] \, ,
\end{equation}
 where the indices \( \mu ,\nu  \) cover the \((4M+2)\)-dimensional
phase-space of the real and imaginary parts of \( \overrightarrow{\alpha } \)
, so that \( \widetilde{\alpha }=(\overrightarrow{x},\overrightarrow{y}) \),
and \( \partial _{\mu }=\partial /\partial \widetilde{\alpha }_{\mu } \).
The diffusion matrix \( \widetilde{\underline{\underline{D}}}=\widetilde{\underline{\underline{B}}}\, \widetilde{\underline{\underline{B}}}^{T} \)
is now positive semi-definite, since, by construction \begin{equation}
\widetilde{\underline{\underline{B}}}=\left[ \begin{array}{cc}
0 & \underline{\underline{B}}^{x}\\
0 & \underline{\underline{B}}^{y}
\end{array}\right] \, \, \, .
\end{equation}

so that the diffusion matrix is the square of a real matrix --- explicitly,

\begin{equation}
\widetilde{\underline{\underline{D}}}=\left[ \begin{array}{cc}
0 & \underline{\underline{B}}^{x}\\
0 & \underline{\underline{B}}^{y}
\end{array}\right]  \left[ \begin{array}{cc}
0 & 0\\
\left( \underline{\underline{B}}^{x}\right) ^{T} & \left( \underline{\underline{B}}^{y}\right) ^{T}
\end{array}\right] \, \, \, .
\end{equation}
 As \( \mathcal{L}_{GA} \) is now explicitly real as well as positive-definite
by construction, it can be applied to the Hermitian conjugate kernel
as well, resulting in the final time-evolution equation,

\begin{eqnarray}
\frac{\partial \widehat{\rho }}{\partial t} & = & \int G(\widetilde{\alpha })\left[ \mathcal{L}_{GA}\widehat{\Lambda }(\widetilde{\alpha })\right] d^{\,4M+2}\widetilde{\alpha }\, \, .
\end{eqnarray}

On integrating by parts, \textit{provided boundary terms vanish},
at least one solution will satisfy the following (normally-ordered)
positive-definite Fokker-Planck equation --- with the differential operators
on the left, each acting on all terms to the right,

\begin{equation}
\frac{\partial G}{\partial t}=\mathcal{L}_{GN}G\equiv \left[ -\partial _{\mu }\widetilde{A}_{\mu }+\frac{1}{2}\partial _{\mu }\partial _{\nu }\widetilde{D}_{\mu \nu }\right] G\, .
\end{equation}

This implies that we have an equivalent set of Ito stochastic differential
equations available, with \( 2M \) \emph{real} Gaussian noises \( dW_{i} \)
, which are\begin{eqnarray}
d\Omega  & = & \Omega (\, Vdt+g_{k}dW_{k}\, )\nonumber \\
d\alpha _{j} & = & (A^{(+)}_{j}-g_{k}B_{jk})dt+B_{jk}dW_{k}\, \, \, \, \, \, .\label{gaugeeq} 
\end{eqnarray}
 The noises obey \( \langle dW_{i}dW_{j}\rangle=\delta _{ij}dt \), and are uncorrelated
between time steps.

Numerical simulations are usually done in the Stratonovich calculus,
due to superior convergence properties \cite{MD-semimp}, so the equivalent
\emph{complex} Stratonovich equation allows us to write efficient
algorithms, \begin{eqnarray}\label{strat}
d\alpha _{a} & = & dx_{a}+idy_{a}\nonumber \\
 & = & \left[ A_{a}-\frac{1}{2}\left( B_{bk}\partial _{b}\right) B_{ak}\right] dt+B_{ak}dW_{k}\, \, ,\label{SDE0} 
\end{eqnarray}
 where \( \left( B_{bk}\partial _{b}\right) \equiv \left( B_{bk}^{x}\partial ^{x}_{b}+B^{y}_{bk}\partial ^{y}_{b}\right)  \).
The derivative terms above are the Stratonovich correction in the
drift, corresponding to related terms obtained in curved-space path
integrals.

These gauge terms are now utilized to stabilize coherent-state paths
entering into highly non-classical regions of phase space. This allows
one to benefit from the over-completeness of coherent states, in reducing
the sampling error and eliminating boundary terms.

\subsection{Moments}

The procedure for calculating observable moments is slightly different
for the gauge representation than for the positive P. Any moment can
be written in terms of the normally ordered operator products \( \hat{a}^{\dagger n}\hat{a}^{m} \),
and their expectation values are given by \begin{equation}
\label{moments}
\left\langle \hat{a}^{\dagger n}\hat{a}^{m}\right\rangle _{\text {quant}}=\frac{\left\langle \beta ^{n}\alpha ^{m}\Omega +(\alpha ^{n}\beta ^{m}\Omega )^{*}\right\rangle _{\text {stoch}}}{\left\langle \Omega +\Omega ^{*}\right\rangle _{\text {stoch}}}\, \, .
\end{equation}
 which differs from the positive-P situation whenever \( \Omega  \)
differs from unity.

The average norm \( \langle \Omega \rangle  \) is always preserved
if there is no potential term (\( V=0 \) ), since the resulting equation
for the weight variable is\begin{equation}
d\Omega =\Omega g_{k}dW_{k}\, \, .
\end{equation}
 The decorrelation property of Ito equations \cite{Arnold} then implies
that \begin{equation}
\langle d\Omega \rangle =\langle \Omega g_{k}\rangle \langle dW_{k}\rangle =0\, \, .
\end{equation}

\subsection{Gauge properties}

We turn briefly here to the question of gauge classification and properties.
Just as in QED, the over-complete nature of the coherent-state expansion
means that many equivalent, stable gauges exist. However, they may
not be equivalent in terms of boundary terms. These are determined
by the tails of the distribution function, which depends intimately
on the gauge chosen for the time evolution. It is essential that the
distribution tails are sufficiently bounded to eliminate boundary
terms arising in partial integration. It is sufficient to bound tails
better than any inverse power law, for which it is conjectured to
require (as a necessary condition) that all deterministic trajectories
are bounded over any finite time interval \cite{GGD-Validity}. This
issue is discussed in greater detail below, and in Ref. \cite{gauge_paper}.

The main criteria for a useful gauge are the elimination of boundary
terms and the reduction of sampling error. However, there is an enlarged
space of variables for the Fokker-Planck equation here. For this reason,
it is possible to stabilize trajectories in the usual positive-P phase space,
while introducing new gauge-induced boundary terms in the \( \Omega  \)
space. When it comes to the formation of boundary terms, the phase of
\( \Omega  \) is generally innocuous provided the gauge is periodic
in this variable, but the gauge distribution must be strongly bounded
as \( |\Omega |\rightarrow \infty  \) to prevent new boundary terms
from arising.

We can classify gauges according to their real or imaginary nature,
and their functional dependence; which can be on just the phase-space
variables, just the quantum phase, or on both. This gives rise to
nine gauge types, depending on the following criteria.

\subsubsection{Gauge complexity}

\label{GCOMPLEX} Gauges are in general complex functions, which leads
to the following classification of gauge complexity:

\begin{enumerate}
\item Real gauge 
\item Imaginary gauge 
\item Complex gauge 
\end{enumerate}
In general, we find that trajectories can be stabilized by real, imaginary
or complex gauges, provided they have some \( ({\bm \alpha },{\bm \beta }) \)
phase-space dependence.

It is worthwhile to note that the imaginary and real parts of the gauges
affect the behavior of sampling error differently. In the Ito calculus,
the evolution of the weight \( \Omega  \) due to the gauges is simply
\( d\Omega =\Omega g_{k}dW_{k} \). Typically, i.e., when there are
no significant correlations between the phase of \( {\bm \alpha } \)
( or \( {\bm \beta } \)) and \( \Omega  \), the weight factor appearing
in moment calculations is just approximately \( \text{Re}[\Omega ] \).
As a general rule, sampling errors are partially due to stochastic
fluctuations in the phase-space trajectories, and partially due to
stochastic fluctuations in the weight function. Thus there is a trade-off;
a gauge that is strongly stabilizing may reduce phase-space fluctuations
at the expense of increased weight variance, and vice versa.

To understand the different types of gauges in somewhat greater detail,
we consider the evolution of the weight variance for real and imaginary
gauges, in a simple case where gauge and weight are decorrelated,
with \( \Omega =1 \) initially. Let \( \Omega =\Omega '+i\Omega '' \)
and \( g_{k}=g'_{k}+ig''_{k} \)~, then \begin{eqnarray}
d\Omega ' & = & \left( \Omega 'g'_{k}-\Omega ''g''_{k}\right) dW_{k}\, \, ,\nonumber \\
d\Omega '' & = & \left( \Omega 'g''_{k}+\Omega ''g'_{k}\right) dW_{k}\, \, .
\end{eqnarray}
If we consider the evolution of the squares of these terms, the Ito
rules of stochastic calculus give \begin{eqnarray}
d\langle [\Omega ']^{2}\rangle  & = & \langle \left( \Omega 'g'_{k}-\Omega ''g''_{k}\right) ^{2}\rangle dt\, \, ,\nonumber \\
d\langle [\Omega '']^{2}\rangle  & = & \langle \left( \Omega 'g''_{k}+\Omega ''g'_{k}\right) ^{2}\rangle dt\, \, .\label{wvariance} 
\end{eqnarray}
 Suppose for simplicity that the \( g_{k} \) and \( \Omega  \) are
approximately uncorrelated, then we have two cases to consider.

\begin{enumerate}
\item Real gauge:\begin{eqnarray}
d\langle [\Omega ']^{2}\rangle  & = & \langle [\Omega ']^{2}\rangle d\tau \, \, ,
\end{eqnarray}
where \( d\tau =\langle g_{k}g_{k}\rangle dt \). This initially leads to
linear
growth in the variance, and hence in the sampling error. The real
part of the gauge will cause noise directly in \( \Omega ' \), producing
asymetric spreading in \( \Omega ' \), which can lead to a few rare
very highly weighted trajectories for times \( \tau \gtrsim 1 \).
The effect of the real gauge may become misleading once the distribution
becomes highly skewed, as the rare trajectories that are important
for moment calculations may be missed if the sample is too small.
At long times, if \( \langle g_{k}g_{k}\rangle  \) is constant and
uncorrelated with \( \Omega  \), then the growth becomes exponential,
with \( \langle [\Omega ']^{2}\rangle =e^{\tau } \).

\item Imaginary gauge:\begin{eqnarray}
d\langle [\Omega ']^{2}\rangle  & = & \langle [\Omega '']^{2}\rangle d\tau \, \, ,\nonumber \\
d\langle [\Omega '']^{2}\rangle  & = & \langle [\Omega ']^{2}\rangle d\tau \, \, .
\end{eqnarray}
 where \( d\tau =\langle g''_{k}g''_{k}\rangle dt \). This leads initially
to  quad\-ratic growth in the variance of $\Omega'$, and hence a slower growth in
the sampling error. If \( \langle g_{k}g_{k}\rangle  \)
is constant and remains uncorrelated with \( \Omega  \), then the
growth is given by \( \langle [\Omega ']^{2}\rangle =\cosh (\tau ) \),
\( \langle [\Omega '']^{2}\rangle =\sinh (\tau ) \). An imaginary
gauge will cause mutual canceling of trajectories that have weights
of randomly positive and negative sign once \( \tau \gtrsim \pi  \). This
can also have deleterious effects for small samples, if the average
sample weight becomes negative --- of course, this cannot be true over
the entire stochastic population.
\end{enumerate}
The generic behavior is more complex than in the examples given above,
due to correlations between the gauge and the normalization. 

Clearly any type of gauge tends to cause growth in the norm variance.
However, there is an exception to this rule: the norm-preserving gauges.
This class of gauges is of special interest as they generate trajectories
having an invariant normalization, so that \( \text{Re} [d\Omega ]\equiv 0 \).
From the equation for the norm variance, Eq. (\ref{wvariance}), it
follows that a necessary and sufficient condition for a norm-preserving
gauge is that \( \Omega 'g'_{k}=\Omega ''g''_{k}\,  \). If \( \Omega '=1 \)
initially, this implies that \( g_{k}=i\Omega ^{*}f_{k}=i(1-i\Omega '')f_{k} \)
, where \( f_{k} \) is a real function. Unless \( g_{k}=0 \), norm-preserving
gauges are generally functions of both the phase-space variables and
the weight \( \Omega  \)~. A preliminary study of these gauges has
shown that these gauges can greatly reduce sampling error, although
gauge-induced boundary terms are also possible \cite{ccp2k}, depending
on the choice of \( f_{k} \).

\subsubsection{Functional dependence}

From the above analysis, we see that gauges can functionally depend
on any phase-space variable, as well as the generalized quantum phase
variable or weight \( \Omega  \)~. This leads to three functional
types:

\begin{enumerate}
\item Autonomous (depends on \( \Omega  \) only) 
\item Space dependent (depends on phase-space only) 
\item Mixed (depends on all components of \( \overrightarrow{\alpha } \)
including \( \Omega  \) ) 
\end{enumerate}
Autonomous gauges appear to be the least useful since they do not affect
\( {\bm \alpha } \) or \( {\bm \beta } \) behavior, but gauges of
either purely space-dependent or mixed type can be used.

A possible caveat with mixed gauges is that they may be much harder
to analyze, as two-way couplings will occur between the normal 
phase-space variables \( {\bm \alpha } \), \( {\bm \beta } \) and the
weight.

\section{Nonlinear Absorber case}

\label{D12PD}

The nonlinear absorber is an example of a nonlinear master equation
that can give either correct or incorrect results when treated with
the usual positive-P representation methods, if the boundary
terms are ignored. Generally, problems only arise when the linear
damping has exceptionally small values or the number of bosons per
mode is small (see Fig.~\ref{2PD-varN}), so this is not a
practical problem in optics. However, for other physical systems such
as a BEC this may be significant. It is a well-studied case, and a
detailed treatment can be found in Ref. \cite{GGD-Validity}. It also has
the merit that exact solutions can be readily found using other means.
By analyzing this example we can ensure that the modifications to
the drift equations obtained from gauge terms, do eliminate boundary
terms and give correct results.

Consider a cavity mode driven by coherent radiation, and damped by
a zero temperature bath that causes both one and two photon losses.
We have scaled time so that the rate of two-photon loss is unity.
Without this nonlinear process, nothing unusual happens. The scaled one-photon loss rate
is \( \gamma \), and \(\varepsilon\) is the scaled (complex) driving field amplitude.  The master
equation is\begin{eqnarray}
\frac{\partial \hat{\rho }}{\partial t} & = & \left[ \varepsilon \hat{a}^{\dagger }-\varepsilon ^{*}\hat{a},\hat{\rho }\right] +\frac{\gamma }{2}(2\hat{a}\hat{\rho }\hat{a}^{\dagger }-\hat{a}^{\dagger }\hat{a}\hat{\rho }-\hat{\rho }\hat{a}^{\dagger }\hat{a})\nonumber \\
 &  & \, \, +\, \, \frac{1}{2}(2\hat{a}^{2}\hat{\rho }\hat{a}^{\dagger 2}-\hat{a}^{\dagger 2}\hat{a}^{2}\hat{\rho }-\hat{\rho }\hat{a}^{\dagger 2}\hat{a}^{2})\, \, .\label{12pdmaster} 
\end{eqnarray}

Following the treatment of Sec.~\ref{GAUGES}, we arrive at the gauge
representation Stratonovich stochastic equations \begin{eqnarray}
d\alpha  & = & [\varepsilon ^{\, }-\alpha (\alpha \beta +ig+(\gamma -1)/2)]dt+i\alpha dW\, \, ,\nonumber \\
d\beta  & = & [\varepsilon ^{*}-\beta (\alpha \beta +i\overline{g}+(\gamma -1)/2)]dt+i\beta d\overline{W}\, \, ,\nonumber \\
d\Omega  & = & S_{\Omega}dt + \Omega \left[ gdW+\overline{g}d\overline{W}\, \, \right] \, \, \, \, .\label{baseeq} 
\end{eqnarray}
 Here \( S_{\Omega }dt \) is the appropriate Stratonovich correction
term [given by the derivative terms in Eq.~(\ref{strat})\,], which depends on the particular gauges chosen.

With no gauge (\( g=\overline{g}=0 \)), the positive-P Stratonovich
equations are recovered,\begin{eqnarray}
d\alpha  & = & [\varepsilon \, \, \, -\alpha (\alpha \beta +\{\gamma -1\}/2)]dt+i\alpha dW\, \, ,\nonumber \label{12pdposp} \\
d\beta  & = & [\varepsilon ^{*}-\beta (\alpha \beta +\{\gamma -1\}/2)]dt+i\beta d\overline{W}\, \, .
\end{eqnarray}

We will concentrate on the various simplifications of this model,
which correspond to existing literature, and simpler analysis.

\subsection{Relevance to many-body problems}

The nonlinearity seen here can occur directly in the form of a nonlinear
collisional damping term in a many-body system, so that it can be
referred to generically as `two-boson absorption'. This type of damping
is common both to nonlinear photonic and atomic interactions.

It is of nearly the same form as for an `imaginary-time' thermal equilibrium
state calculation for the usual model of an alkali-metal Bose gas
or BEC \cite{BECH}. There, for example, the interaction energy between
identical bosons of mass \( m \) and \( s \)-wave scattering length
\( a_{s} \) in \( D \)-dimensional space is given by \begin{equation}
\hat{H}=\frac{2\pi \hbar ^{2}a_{s}}{m}\int d^{D}\mathbf{x}\hat{\psi }^{\dagger2 }(\mathbf{x})\hat{\psi }^{2}(\mathbf{x})\, \, ,
\end{equation}
 provided that \( a_{s} \) is much smaller than other characteristic
lengths of the system (which is usually the case). The master equation
for an imaginary-time calculation is \begin{equation}
\frac{\partial \hat{\rho }_{e}}{\partial \tau }=-\frac{1}{2}\left\{ \hat{H}-\mu \hat{N},\hat{\rho }_{e}\right\} _{+}\, \, ,
\end{equation}
 where \( \hat{\rho }_{e} \) is the thermal canonical ensemble density
matrix, \( \mu  \) is the chemical potential, \( N \) is the number operator
for the entire system, and \( \tau =1/k_{B}T \) is an inverse temperature.
Apart from the fact that it is not trace-preserving, this is a nonlinearity very
similar to that occurring in the nonlinear absorber master equation.

While boundary-term discrepancies only occur with this nonlinearity
for low occupations per mode (see also Fig.~\ref{2PD-varN}), for
a many-mode system at finite temperature one expects a large number
of modes to have just such a low occupation. Thus, it is important
to check that boundary terms are indeed eliminated. Note that the
gauge representation simulation is efficient over a wide range of
occupation numbers. See, for example, Fig.~\ref{12PD}. More details
of applications to both real and imaginary time many-body systems
with many modes will be given elsewhere.

\subsection{Two-boson absorber}

\label{2PD} In its simplest form, corresponding to \( \gamma =\varepsilon =0 \),
only two-boson absorption takes place. We expect that for a state
\( \left| \psi \right\rangle =\sum _{n}c_{n}|n\rangle  \) all even
boson number components will decay to vacuum, and all odd-numbered
components will decay to \( \left| 1\right\rangle  \), leaving a
mixture of vacuum and one-boson states at long times.

The positive-P representation has been found to give erroneous results \cite{SG-abs,CM-abs,MC-abs,GGD-using}
due to the existence of moving singularities \cite{GGD-Validity},
which cause power-law tails in the distribution leading to boundary
terms. The moment usually concentrated on in this system is the number
of bosons \( \hat{n}=\hat{a}^{\dagger }\hat{a} \), which corresponds
to the statistical average of \( n=\alpha \beta  \) in the positive-P representation. 
This has a convenient closed equation (Stratonovich),
\begin{equation}
\label{2PDeq}
dn=-n(n+i\tilde{g}-1/2)d\tau +indW^{+}
\end{equation}
 with \( dW^{+}=(dW+d\overline{W}) \), \( \tau =2t \), and \( \tilde{g}=(g+\overline{g})/2 \).

Let us examine the behavior of the above equation, when \( \tilde{g}=0 \),
i.e., in the standard, un-gauged formulation. The deterministic part
of the evolution has a repellor at \( n=0 \), and an attractor at
\( n=\frac{1}{2} \). The noise is finite, and of standard deviation
\( \sqrt{dt/2} \) at the attractor. We can see that the deterministic
part of the evolution has a single trajectory of measure zero which
can escape to infinity along the negative real axis, \begin{equation}
\alpha =-\beta =\frac{1}{\sqrt{\tau _{0}-\tau}},
\end{equation}
 where \( \tau _{0}=1/\alpha (0)^{2}=-1/n(0) \). This moving singularity
is known to cause the power law behavior of the Fokker-Planck solution
at large \( |n| \), which means that integration by parts is not
in fact valid --- which leads to incorrect results.

Indeed, it can be easily seen that in the steady-state limit, all
trajectories in a simulation will head toward \( n=\frac{1}{2} \),
making \(\lim_{t\rightarrow\infty} \langle \hat{n}\rangle =\frac{1}{2} \).
Quantum mechanics, however, predicts that if we start from a state
\( \hat{\rho }_{0} \), the steady state will be \begin{equation}
\lim _{t\rightarrow \infty }\langle \hat{n}\rangle =\sum _{j=0}^{\infty }\left\langle 1+2j\right| \hat{\rho }_{0}\left| 1+2j\right\rangle \, \, .
\end{equation}
 For a coherent state \( \left| \alpha _{0}\right\rangle  \) input,
say, this will be \begin{equation}
\label{exss}
\lim _{t\rightarrow \infty }\langle \hat{n}\rangle =\frac{1}{2}\left( 1-e^{-2|\alpha_{0} |^{2}}\right) \, \, .
\end{equation}
 Thus we can expect that the positive P simulation will give correct
results only when \( e^{|\alpha_{0} |^{2}}\gg 1 \).

To correct the problem we have to change the phase-space topology
in some way to prevent the occurrence of moving singularities. We
have found that a good gauge for a two-boson absorber nonlinearity
in general is \begin{equation}
\label{goodgauge}
g=\overline{g}=\tilde{g}=i(n-|n|)\, \, .
\end{equation}
 This replaces the \( -n^{2} \) term in Eq. (\ref{2PDeq}), which may
become repulsive from zero, with \( -n|n| \) which is always a restoring
force, and so never leads to super-exponential escape.

With the gauge (\ref{goodgauge}), the Stratonovich equations become
\begin{eqnarray}\label{2pdeqs}
dn & = & -n(|n|-1/2)d\tau +indW^{+}\, \, ,\\
d\Omega  & = & \Omega \left\{\ [n+(n-|n|)^2]d\tau/2+i(n-|n|)dW^{+}\, \, \right\} \, \, .\nonumber 
\end{eqnarray}
 Phase-space trajectories have changed now, but since it has all come
from the same master equation, it still describes the same system.
Consider the equations for the polar decomposition of \( n=re^{i\phi } \),
\begin{eqnarray}
dr & = & -r(r-1/2)d\tau \, \, ,\nonumber \\
d\phi  & = & dW^{+}\, \, .
\end{eqnarray}
 This is exact, and shows that now we have an attractor on the circle
\( |n|=\frac{1}{2} \), and a repellor at \( n=0 \), with free phase
diffusion in the tangential direction. Once trajectories reach the
attractor, only phase diffusion occurs. Some more complicated evolution
is occurring in the \( \Omega  \) variable. In any case, there are
now no moving singularities anywhere in the phase space, and simulations
correspond exactly to quantum mechanics.

\begin{figure}
\includegraphics*[width=8.6cm]{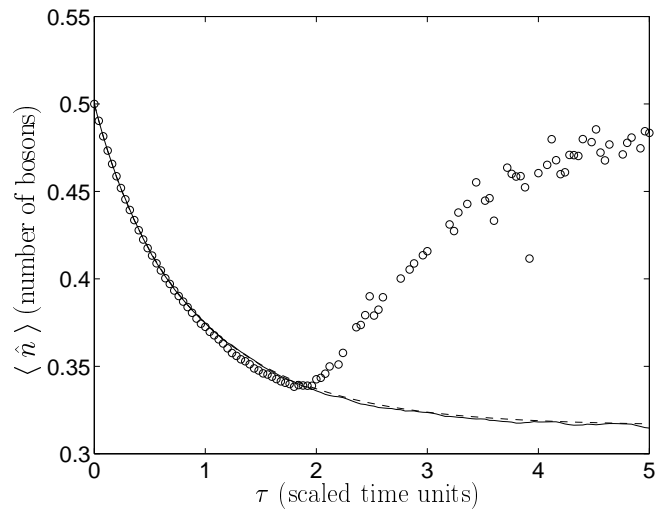}
\caption{Comparison of two-boson damping simulations. \textit{Circles}:
positive P simulation; \textit{solid line}: circular gauge simulation;
\textit{dashed line}: exact calculation (truncated number-state basis).
Simulation parameters: 40 000 trajectories; step size = 0.005; initial
coherent state. Stratonovich semi-implicit method \cite{MD-semimp}. 
\label{2PD-exact+posp+gauge}}
\end{figure}

\begin{figure}
\includegraphics*[width=8.6cm]{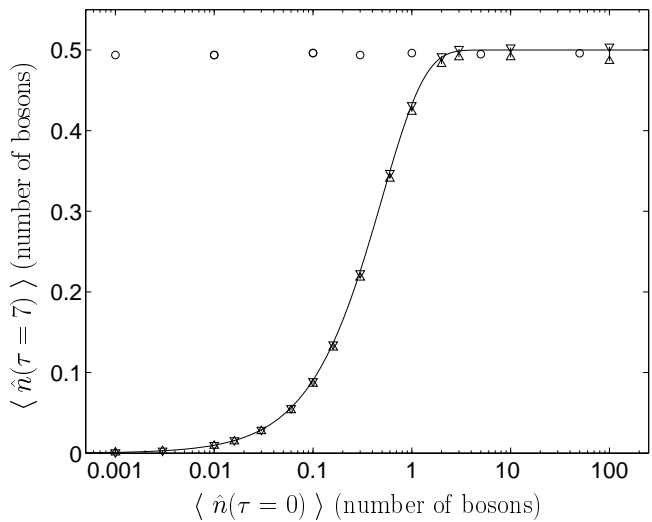}
\caption{Steady state expectation values of boson number \protect\(\langle\hat{n}\rangle\protect \)
obtained by gauge simulations (\textit{double triangles}) compared
to exact analytic results from Eq. (\ref{exss})\ (\textit{solid line})
and positive-P simulations (\textit{circles}) for a wide range of
initial coherent states. Size of uncertainty in gauge results due
to finite sample size is indicated by vertical extent of `double-triangle'
symbol. Steady state was observed to have been reached in all simulations
by \protect\( \tau =7\protect \) or earlier (compare with Fig.~\ref{2PD-exact+posp+gauge}
and \ref{12PD}), hence this is the time for which the simulation
data is plotted. Simulation parameters: 100 000 trajectories; step size
= 0.01. \label{2PD-varN}}
\end{figure}

Figure \ref{2PD-exact+posp+gauge} compares results for a truncated
number-state basis calculation, a positive-P calculation, and a {}``circular''
gauge (\ref{goodgauge}) calculation for an initial coherent state
of \( \alpha _{0}=1/\sqrt{2} \). Figure~\ref{2PD-varN} compares steady-state
 values for exact, positive-P, and gauge calculations for various
initial coherent states in a wide range. It is seen that the gauge
calculation is correct to within the small errors due to finite sample
size.

\subsection{One- and two-boson absorber}

If we now turn on the one-boson decay as well, but still do not have
any driving, we expect that all states will decay to the vacuum on
two time scales \( 1 \) and \( 1/\gamma  \). If \( \gamma \gg 1 \),
nothing interesting happens, however if \( \gamma \lesssim 1 \),
we should first see a rapid decay to a mixture of vacuum and one-boson
states due to the two-boson process, and then a slow decay of the
one-boson state to the vacuum on a time scale of \( \tau \approx 2/\gamma  \).

In this case the positive-P equations display different behavior depending
on whether \( \gamma  \) is above or below the threshold \( \gamma =1 \).
Below threshold, we have an attractor at \( n=(1-\gamma )/2 \), and
a repellor at \( n=0 \), while above threshold, the attractor is
at \( n=0 \), and the repellor at \( n=-(\gamma -1)/2 \). In either
case, there is a singular trajectory along the negative real axis,
which can cause boundary-term errors. It turns out that the steady
state calculated this way is erroneous while \( \gamma <1 \), and
there are transient boundary-term errors while \( \gamma <2 \) \cite{SG-abs}.
The false steady state below threshold lies at the location of the
attractor: \( (1-\gamma )/2 \).

Let us try to fix this problem using the same circular gauge (\ref{goodgauge})
as before. The equation for \( r \) is now \begin{equation}
dr=-r(r-[1-\gamma ]/2)d\tau \, \, ,
\end{equation}
 while the \( \phi  \) and \( \Omega  \) evolution is unchanged.
So, above threshold we are left with only an attractor at \( n=0 \),
while below threshold we have a repellor at \( n=0 \) surrounded
by an attracting circle at \( r=(1-\gamma )/2 \). This phase space
again has no moving singularities.

\begin{figure}
\includegraphics*[width=8.6cm]{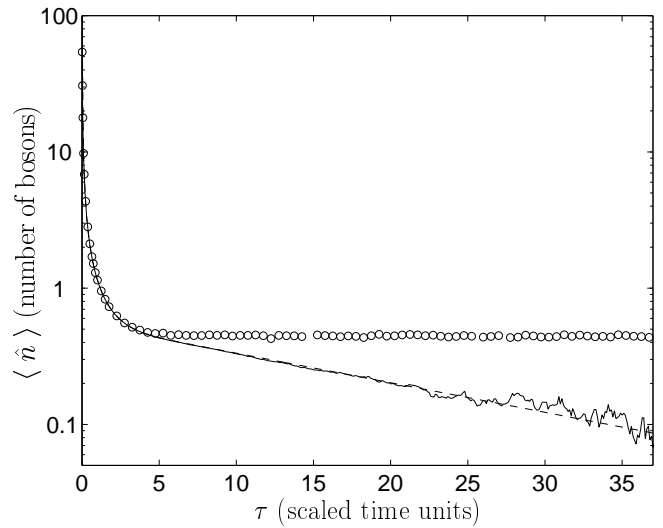}
\caption{Comparison of simulations for system with both single- and 
double-boson damping. Relative strength \protect\( \gamma =0.1\protect \);
\textit{Circles}: positive-P simulation. \textit{solid line}: circular
gauge simulation; \textit{dashed line}: exact calculation (truncated
number-state basis). Gauge simulation parameters: \protect\( 10^{5}\protect \)
trajectories; step size varies from \protect\( 0.0001\protect \)
to \protect\( \approx 0.006\protect \); initial coherent state \protect\( \left| 10\right\rangle \protect \)
with \protect\( \langle \hat{n}\rangle =100\protect \) bosons. 
\label{12PD}}
\end{figure}

The results of simulations for the parameter \( \gamma =0.1 \) are
shown in Fig.~\ref{12PD}. The gauge simulation tracks the exact
results. We have chosen \( \gamma \ll 1 \) so that a system with
two widely differing time scales is tested. The circular gauge avoids
the false results of the positive-P simulation. Note also that the
gauge simulation remains efficient for a wide range of occupation
numbers --- from \( \langle \hat{n}\rangle \approx 100\gg 1 \), where
the positive P is also accurate, to \( \langle \hat{n}\rangle \approx 0.1\ll 1 \)
where it is totally incorrect.

\subsection{Driven two-boson absorber}

\begin{figure}
\includegraphics*[width=8.6cm]{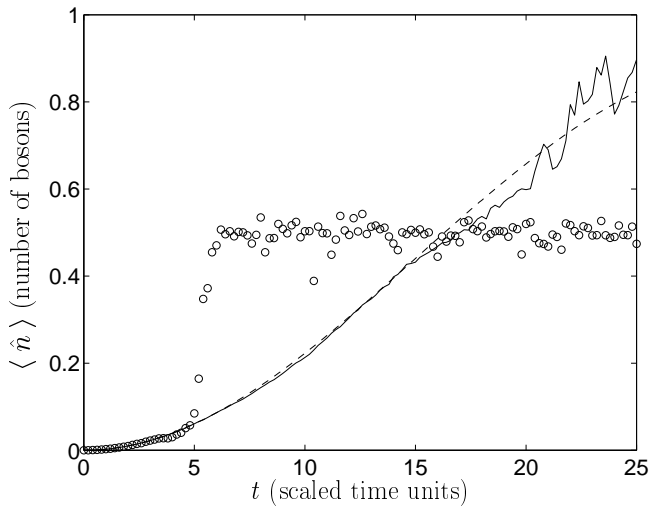}
\caption{Driven two-boson absorber with \protect\( \varepsilon =0.05\protect \).
\textit{Circles}: positive P simulation (\protect\( 1000\protect \)
trajectories); \textit{solid line}: circular gauge simulation (\protect\( 10^{5}\protect \)
trajectories); \textit{dashed line}: exact calculation (truncated
number-state basis). Step size \protect\( \Delta t=0.025\protect \).
Initial vacuum state. 
\label{TPDE}}
\end{figure}

The other type of situation to consider is when we have a driving
field as well as two-boson damping. In these considerations we have
set the one-boson damping rate to zero (\( \gamma =0 \)), since this
process never causes any of the simulation problems anyway, but leaving
it out simplifies analysis. Failure of the positive-P representation
method has been found in this limit as well  \cite{SS-fail}, and is
evident in Fig.~\ref{TPDE}. The equation for \( n \) is no longer
stand-alone in this case, and we must simulate all three complex variables
as in Eq.~(\ref{baseeq}), the $\Omega$ equation being the same as in the 
undriven case (\ref{2pdeqs}).

A treatment of the singular trajectory problem with the same circular
gauge~(\ref{goodgauge}) leads again to correct results, as seen
in Fig.~\ref{TPDE}.

\section{The single-mode laser}

\label{LAS} Let us now consider the second quantum system for which
systematic errors have been seen with the positive-P representation.
We will see that the problem here is somewhat different than in the
previous case. The difference is that for two-boson damping, boundary-term
errors occur even when we choose an optimal (i.e., compact) initial
distribution to represent our starting state, whereas here systematic
errors occur only for unreasonably broad initial distributions. Nevertheless,
since normally it is assumed that the initial condition can be of
arbitrary breadth it is instructive to investigate how this problem
can be tackled with stochastic gauge methods.

We have found that stochastic gauges can be used to increase the allowable
breadth to include all reasonable starting conditions, but once one
tries to increase the initial spread too much, it becomes unlikely
that any gauge will remove systematic errors, without introducing
too much sampling (i.e. random) error instead.

\subsection{The laser model}

Ito stochastic differential equations for a simple photonic or atomic
laser model that can be derived from the positive-P distribution are \cite{SS-fail,GGD-Validity}\begin{eqnarray}
d\tilde{\alpha } & = & (G-\tilde{\alpha }\tilde{\beta })\tilde{\alpha }d\tau +\sqrt{Q}d\eta \, \, ,\nonumber \\
d\tilde{\beta } & = & (G-\tilde{\alpha }\tilde{\beta })\tilde{\beta }d\tau +\sqrt{Q}d\eta ^{*}\label{sseqns} 
\end{eqnarray}
 in appropriate scaled variables, with the complex Gaussian noise
\( d\eta  \) obeying \( \langle d\eta d\eta ^{*}\rangle =2d\tau  \).
In terms of physical parameters, we have \begin{eqnarray}
\tilde{\alpha } & = & \alpha /\sqrt{\mathcal{N}}\nonumber \\
\tilde{\beta } & = & \beta /\sqrt{\mathcal{N}}\, \, ,
\end{eqnarray}
 where \( \tau  \) is the scaled time, 
and \( {\mathcal{N}}\gg 1 \) is a scaling parameter
that equals the number of gain atoms in a simple photonic laser model.
Both \( G \) the gain parameter and \( Q\ge G/\mathcal{N} \), the
noise parameter, are real and positive.

Since this time we are again interested in the (scaled) boson number
\( \langle\tilde{n}\rangle = \langle\tilde{\alpha}\tilde{\beta}\rangle = \langle\hat{n}\rangle\!/{\mathcal N} \),
its evolution can
be written as a closed equation \begin{equation}
d\tilde{n}=-2(\tilde{n}-a)(\tilde{n}-b)d\tau +2\sqrt{Q\tilde{n}}dW\, \, ,
\end{equation}
 where now the \textit{real} Gaussian noise obeys \( \langle  dW\,dW \rangle =d\tau  \),
and the deterministic stationary points in the Stratonovich calculus
are \begin{eqnarray}
a & = & \frac{1}{2}\left( G+\sqrt{G^{2}+2Q}\right) \, \, ,\nonumber \\
b & = & \frac{1}{2}\left( G-\sqrt{G^{2}+2Q}\right) \, \, .
\end{eqnarray}
 We find that the stationary point at \( a \) is an attractor, and
at \( b \) we have a repellor. Defining \( \Delta =b-\tilde{n} \),
we get \begin{equation}
d\Delta =2\Delta (\Delta +\sqrt{G^{2}+2Q})+\text {noise}\, \, ,
\end{equation}
 which shows that we again have a singular trajectory escaping to
infinity in finite time along the negative real axis for \( \tilde{n}<b \).

\subsection{Initial conditions}

Let us consider the usual case of vacuum initial conditions. A vacuum
can be represented by \begin{equation}
\label{deltap}
P^{(+)}(\tilde{\alpha },\tilde{\beta })=\delta (\tilde{\alpha })\delta (\tilde{\beta })\, \, ,
\end{equation}
 but also by Gaussian distributions of any variance \( \sigma _{0}^{2} \),
around the above, \begin{equation}
\label{varini}
P^{(+)}(\tilde{\alpha },\tilde{\beta })=\frac{1}{4\pi ^{2}\sigma _{0}^{4}}\exp \left\{ -\frac{|\tilde{\alpha }|^{2}+|\tilde{\beta }|^{2}}{2\sigma _{0}^{2}}\right\} \, \, .
\end{equation}
 Note: the distribution of \( \tilde{n} \) is non-Gaussian, but has
a standard deviation of \( \sigma _{\tilde{n}}\approx \sqrt{2}\sigma _{0}^{2} \)
in both the real and imaginary directions.

It has been found by Schack and Schenzle \cite{SS-fail} that for the
single-mode laser model, a positive-P simulation of pumping from a
vacuum will give correct answers if the usual $\delta$-function initial
condition (\ref{deltap}) is used, but will have systematic errors
if the initial condition used has a sufficiently large variance
(see Fig.~\ref{OML}). We emphasize here that this is not a real
problem in practical cases, as the variance required to cause systematic
errors is typically extremely large, once the scaling needed to obtain
the usual (approximate) laser model is taken into account.

This can be understood because if we have a sufficiently broad initial
distribution, the region of phase space that includes the singular
trajectory will be explored by the distribution. Even if initially
\( \sigma _{\tilde{n}}\ll |b| \), the region \( \tilde{n}<b \) may
be subsequently explored due to the presence of the noise terms.

Apart from the obvious $\delta$-function initial condition, one might
want to try the canonical distribution of Eq.~(\ref{CanonicalP}),
which is a standard positive-P representation construction \cite{DG-PosP}.
It will not cause problems as its variance is \( \sigma _{0}^{2}=1/{\mathcal{N}} \),
which for any realistic case will be very small (i.e., \( \sigma _{\tilde{n}}\ll |b| \)).
Schack and Schenzle discovered anomalous results when they chose \( \sigma _{0}^{2}=1 \),
due to an erroneous procedure of scaling the equations --- while not
scaling the canonical initial condition in \( \alpha  \). Nevertheless,
since any \( \sigma _{0} \) is supposed to represent the same state,
insight into what can be achieved using gauge methods is gained if
we analyze the systematic errors for such a relatively large \( \sigma _{0} \).

\subsection{Gauge corrections}

The Fokker-Planck equation corresponding to Eq.~(\ref{sseqns}) is \begin{equation}
\frac{\partial P}{\partial \tau }=\left\{ \frac{\partial }{\partial \tilde{\alpha }}[\tilde{n}-G]\tilde{\alpha }+\frac{\partial }{\partial \tilde{\beta }}[\tilde{n}-G]\tilde{\beta }+2Q\frac{\partial ^{2}}{\partial \tilde{\alpha }\partial \tilde{\beta }}\right\} P\, \, .
\end{equation}
 We now introduce gauges using the same method as in Sec.~\ref{GAUGES}.
This leads to the Ito stochastic equations \begin{eqnarray}
\textstyle d\tilde{\alpha } & = & \tilde{\alpha }(G-\tilde{n})d\tau -\sqrt{Q}(g+i\overline{g})d\tau +\sqrt{Q}d\eta \, \, ,\nonumber \\
d\tilde{\beta } & = & \tilde{\beta }(G-\tilde{n})d\tau -\sqrt{Q}(g-i\overline{g})d\tau +\sqrt{Q}d\eta ^{*} \, \, ,\nonumber \\
d\Omega  & = & \Omega \, [\, \, (g-i\overline{g})d\eta +(g+i\overline{g})d\eta ^{*}\, ]/2\, \, .
\end{eqnarray}
 It is convenient to define a transformed gauge function \( \tilde{g} \),
which is also arbitrary, such that \begin{eqnarray}
g=\frac{(\tilde{\alpha }+\tilde{\beta })\tilde{g}}{2\sqrt{Q}} &  & \, \, ,\nonumber \\
\overline{g}=\frac{(\tilde{\alpha }-\tilde{\beta })\tilde{g}}{2i\sqrt{Q}}\, \, .
\end{eqnarray}
 Changing to \( \tilde{n} \) and \( \Theta =\ln (\Omega ) \) variables
we obtain the Stratonovich equation \begin{eqnarray}
d\tilde{n} & = & 2\tilde{n}(G-\tilde{n}-\tilde{g})d\tau +Qd\tau +2\sqrt{Q\tilde{n}}\, dW\, \, ,\nonumber \\
d\Theta  & = & -\frac{\tilde{n}\tilde{g}^{2}}{2Q}d\tau +S_{\Theta }d\tau +\tilde{g}\sqrt{\frac{\tilde{n}}{Q}}\, dW\, \, ,
\end{eqnarray}
 with \( S_{\Theta }dt \) being the appropriate
Stratonovich correction [given by the derivative terms in Eq.~(\ref{strat})\,] for a particular gauge function \( \tilde{g} \).

\subsection{Correcting for the moving singularities}

Consider the deterministic evolution of the real part, \( \tilde{n}_{x} \),
of \( \tilde{n}=\tilde{n}_{x}+i\tilde{n}_{y} \), \begin{equation}
d\tilde{n}_{x}=-2\tilde{n}_{x}^{2}+2G\tilde{n}_{x}+Q+2\tilde{n}_{y}^{2}-2\tilde{n}_{x}\text{Re} [\tilde{g}]+2\tilde{n}_{y}\text{Im} [\tilde{g}]\, \, .
\end{equation}
 The moving singularity is due to the \( -2\tilde{n}_{x}^{2} \) leading
term for negative values of \( \tilde{n}_{x} \). We now consider
criteria for choosing the drift gauges as follows.

(1) It is desirable to keep the gauge terms to a minimum because whenever
they act the weights of trajectories become more randomized --- see
Sec.~\ref{GCOMPLEX}. Thus, let us restrict ourselves to functions
\( \tilde{g} \) that are only nonzero for \( \tilde{n}_{x}<0 \).

(2) This immediately leads to another restriction on \( \tilde{g} \):
To be able to use the efficient numerical algorithms in the Stratonovich
calculus, we must be able to calculate the correction term \( S_{\Theta } \),
which depends on derivatives of \( \tilde{g}\sqrt{\tilde{n}/Q} \).
This immediately suggests that \( \tilde{g} \) must always be continuous,
hence, in particular,  \( \lim _{n_{x}\rightarrow0 }(\ \tilde{g}\ )=0 \).
For ease of analysis, let us start with a simple form for the gauge,
\( \tilde{g}=c-\lambda \tilde{n}_{x}+\lambda _{y}\tilde{n}_{y} \).
This restriction immediately implies \( c=\lambda _{y}=0 \), hence\begin{equation}
\tilde{g}=\left\{ \begin{array}{cl}
-\lambda \text{Re} [\tilde{n}] & \text {if } \  \text{Re} [\tilde{n}]<0\\
0 & \text {if } \  \text{Re} [\tilde{n}]\geq 0
\end{array}\, \, ,\right. 
\end{equation}
and \( S_{\Theta} = \lambda( \text{Re} [\tilde{n}] + \tilde{n} + |\tilde{n}| )/2. \) when \(\text{Re} [\tilde{n}] <0\), zero otherwise. 

(3) The next necessary condition, to remove moving singularities, is that the \( -2\tilde{n}_{x}^{2} \)
term is canceled, hence: \begin{equation}
\lambda \geq 1\, .
\end{equation}

(4) Now, if \( \lambda =1 \) there are no systematic errors, but the
sampling error very quickly obscures everything because \( n_{x} \)
still heads to \( -\infty  \) exponentially due to the \( 2G\tilde{n}_{x} \)
term. This takes it into regions of everincreasing \( |\tilde{g}| \),
and weights quickly become randomized. For slightly larger parameters
\( \lambda  \), the \( \tilde{n}_{x} \) evolution takes trajectories
to a point lying far into the negative \( n_{x} \) region where the
two leading terms balance. Here the trajectories sit, and quickly
accumulate weight noise. It is clear that for an optimum simulation
all stationary points of \( \tilde{n}_{x} \) in the nonzero gauge
region must be removed. In this system this condition is \begin{equation}
\lambda >1+\frac{G^{2}}{2Q}\, \, .
\end{equation}

\begin{figure}
\includegraphics*[width=8.6cm]{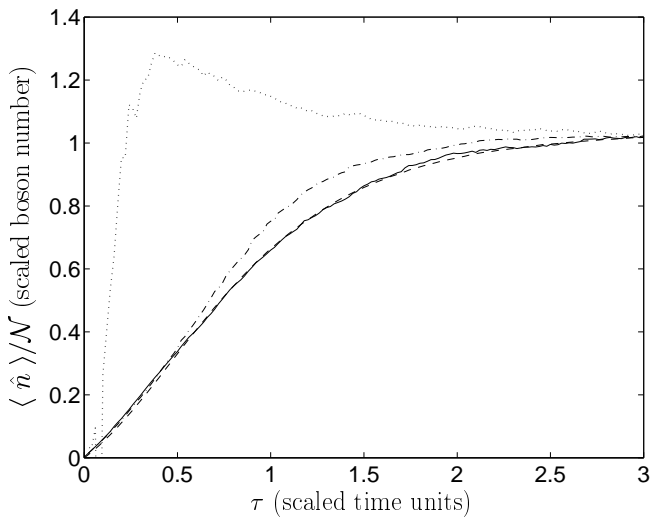}
\caption{One-mode laser \protect\( G=1\protect \), \protect\( Q=0.25\protect \).
\textit{Dashed line}: (correct) positive P simulation with $delta$-function
initial conditions (\ref{deltap}) \protect\( \sigma _{0}^{2}=0\protect \)
and \protect\( 10^{5}\protect \) trajectories. \textit{Dotted-dashed
line}: erroneous positive-P simulation with Gaussian initial conditions
(\ref{varini}) \protect\( \sigma _{0}^{2}=0.1\protect \) initially,
and \protect\( 10^{5}\protect \) trajectories. \textit{Dotted line}:
positive-P simulation with \protect\( \sigma _{0}^{2}=1\protect \),
and \protect\( 10^{4}\protect \) trajectories. \textit{Solid line}:
gauge calculation for \protect\( \sigma _{0}^{2}=0.1\protect \) with
\protect\( \lambda =4\protect \), which corrects the systematic error
of the positive P. Only \protect\( 4000\protect \) trajectories,
so as not to obscure other data. Step size in all cases is 0.005. 
\label{OML}}
\end{figure}

An example has been plotted in Fig.~\ref{OML} where we have parameters
\( G=1 \), \( Q=0.25 \) (leading to \( a\approx 1.1124 \) and \( b\approx -0.1124 \)
). We are considering an initial condition of \( \sigma _{0}^{2}=0.1 \),
which is already much larger than the canonical variance for physically
likely parameters. Typical values of \( \tilde{n} \) initially will
be of order \( \sigma _{\tilde{n}}\approx 0.14\gtrsim |b| \) here.
A good choice of gauge has \( \lambda =4 \). The use of this gauge
clearly restores the correct results.

\subsection{Non-optimal initial conditions}

\label{GSIZE}

As we increase the spread of the initial distribution beyond \( \sigma _{\tilde{n}}\approx |b| \),
it becomes increasingly difficult to find a gauge that will give reasonable
simulations. (For example we have tried a wide variety of what seemed
like promising gauges for \( \sigma_0^{2}=0.3 \), with the above values
of parameters \( Q \) and \( G \), and none have come close to success).
The problem is that while we can remove systematic errors, large random
noise appears and obscures whatever we are trying to calculate.

Trajectories that start off at a value of \( \tilde{n} \) lying
significantly beyond \( b \) require a lot of modification to their
subsequent evolution to (1) stop them from escaping to \( -\infty  \)
and (2) move them out of the gauged region of phase space so that they
do not accumulate excessive weight noise. If there are many of these,
the trade-off between the gauge size and length of time spent in the gauged
region does not give much benefit anymore. Nevertheless, one may be
sure that if this is the case, results will at worst be noisy and
unusable, rather than being systematically incorrect.

We stress again that this whole matter of non-optimal initial
conditions is not a major hurdle to dynamical simulations because
a compact starting distribution is generally found very easily.

\section{Conclusions}

\label{CONCLUSIONS} The positive-P representation is well suited
to complex quantum mechanical problems, such as many-body systems,
but has been known for about a decade to have systematic errors
in some cases of its use --- due to non-vanishing boundary terms. The
gauge P representation, a variant on the usual positive-P representation,
can be used to eliminate boundary terms and consequently all the systematic
errors that were encountered previously. It can also reduce sampling
error in a simulation, and allows `imaginary time' calculations of
thermal equilibrium states. The fact that correct results are immediately
obtained in every case where systematic errors were found with the
positive-P method, is strong evidence that these previous problems
were indeed due to boundary terms caused by moving singularities in
the analytically continued deterministic equations. Of course, boundary
terms can occur for other reasons (for example, if the noise term
grows too rapidly with radius), so caution is still needed in the
gauge choice.

The technique appears to be broadly applicable, and only requires
the recognition of what instabilities in the stochastic equations
could lead to problems. It does not require detailed knowledge of
what the boundary terms are, provided instabilities are removed. However,
we remark here that the general specification of necessary and sufficient
conditions to eliminate boundary terms remains an open problem, and
clearly requires growth restrictions on the gauge terms, both in phase space
and quantum-amplitude space. Care is also required with the choice
of the gauge and initial distribution. However, using unsuitable
gauges or initial conditions may only lead to large sampling errors,
not systematic errors, provided the gauge is chosen to eliminate boundary
corrections in the first place. Sampling error then allows for a confident
assessment of the magnitude of inaccuracies in a simulation, which
can be supplemented by numerical analysis of the distribution tails.

The main conclusion we come to is that this method does, in the cases
studied, provide a complete solution to the problem of simulation
of a many-body quantum system in phase space, under conditions where
previous direct simulation techniques were not practicable. All known
technical requirements on the path to obtaining a stochastically equivalent
description to quantum mechanics, which is applicable to both large
and small particle numbers, have been satisfied by this method. For
this reason, we believe that gauge simulations can be used to simulate
many quantum systems without systematic errors when carrying out more
difficult calculations, where no exact result is known.

These conclusions must be supplemented by the detailed study of relevant
gauges for particular quantum systems. We note, however, that the
mathematical techniques employed here for generating stochastic gauges,
may well be useful for other representations as well as the gauge
P representation described here.

\begin{acknowledgments}
Numerical calculations were carried out using open software from the
XMDS project \cite{xmds}. Thanks to the Australian Research Council
and the Alexander von Humboldt-Stiftung for providing research support.
We thank Damian Pope for suggesting the name stochastic gauges.
\end{acknowledgments}

\appendix

\section{Other extensions of the positive-P representation}

\subsection{The work of Carusotto, Castin, and Dalibard}

Recently, Carusotto, Castin, and Dalibard \cite{Paris1,Paris2} (CCD)
have made related extensions to the positive-P representation. These
were derived for the particular case of an interacting scalar Bose
gas, and led to a number of conditions for an Ito stochastic evolution
to be equivalent to a master equation evolution.

It can be shown quite simply that the equations (\ref{gaugeeq}) generated
by the gauge P representation for this Hamiltonian satisfy the CCD
conditions. We conjecture that these provide the most general possible
solution to the stochastic problem posed by these authors. In particular,
\( db=\Pi[ g_k dW_k - \bar{N}( \phi_1 dB_2^* + \phi_2^* dB_1)] \), using the above paper's formalism. Our methods
can also treat a much larger class of Hamiltonians and master equations
than considered in the CCD treatment.

In Ref.~\cite{Paris1} systematic errors due to boundary terms were not considered. 
However, evolutions satisfying "exactness" conditions derived using the same 
procedure can contain such errors.

As an example, following the CCD procedure  \cite{Paris1} for a one-mode
two-boson absorber master equation, as in Eq. (\ref{12pdmaster})
with \( \gamma =\varepsilon =0 \), one arrives at the conditions \begin{eqnarray}
dB_{1}dB_{2}^{*} & = & 0\nonumber \, \, , \\
dB_{a}^{*2} & = & -\phi _{a}^{2} \, \, ,\nonumber \\
F_{1} & = & -dbdB_{1}/\Pi \, \, , \\
F_{2} & = & -db^{*}dB_{2}/\Pi ^{*} \, \, ,\nonumber \\
f & = & \Pi(\bar{N}\phi_1\phi_2^*)^2 \, \, ,
\end{eqnarray}
 where (referring back to the notation in this present paper), \begin{eqnarray}
d\phi_1 =& d\alpha  /\sqrt{\bar{N}} = & F_1 dt + dB_1 \, \, ,\nonumber \\
d\phi_2 =& d\beta^* /\sqrt{\bar{N}} = & F_2 dt + dB_2 \, \, ,\nonumber \\
d\Pi =& d[ \Omega e^{ -\phi_1\phi_2^* \bar{N} } ] = & f dt + db \, \, .
\end{eqnarray}
 It can be seen that the positive-P equations (\ref{12pdposp}) satisfy
these conditions, while producing the erroneous evolution seen in
Fig.~\ref{2PD-exact+posp+gauge}. In summary, the methods of the
CCD paper do not obviate the need to choose gauges that eliminate
boundary terms.

\subsection{Noise optimization by Plimak, Olsen, and Collett}

\label{PLIMAKA} In  Ref.~\cite{Plimak}, Plimak, Olsen and Collett have
found that for some systems (the Kerr oscillator \( \hat{H}=\omega _{0}\hat{a}^{\dagger }\hat{a}+\kappa \hat{a}^{\dagger 2}\hat{a}^{2}/2 \),
in particular), the most obvious (diagonal) choice of noise matrix
\( B \) may not be the optimal one.

For example, for the above Hamiltonian, one finds that the diffusion
matrix (in \( \alpha ,\beta  \)) variables is \begin{equation}
D=i\kappa \left[ \begin{array}{ccc}
-\alpha ^{2} &  & 0\\
0 &  & \beta ^{2}
\end{array}\right] =BB^{T}\, \, .
\end{equation}
 Following the procedure in Eq. (\ref{transform-B}), an equivalent
but broader choice of noise matrix \( B \) can be any of \begin{equation}
B=\sqrt{i\kappa }\left[ \begin{array}{ccc}
i\alpha \cos (g) &  & i\alpha \sin (g)\\
-\beta \sin (g) &  & \beta \cos (g)
\end{array}\right] \, \, , 
\end{equation}
 with the usual diagonal decomposition given by \( g=0 \). 

However, in Ref.~\cite{Plimak} it was found that for a positive-P simulation,
different decompositions with nonzero constant \( g \) gave the lowest
sampling error for coherent state initial conditions. In their notation,
they introduce \( \sqrt{A+1}=-\sqrt{2}\cos(g) \), and consider the case of
real \( A \ge 1\) (i.e., imaginary \( g \) ) only.

\subsection{Stochastic gauges for the Kerr oscillator}

In Ref.~\cite{ccp2k}, the sampling error in a Kerr oscillator simulation
--- equivalent to a one-mode BEC model, apart from linear terms --- was
reduced substantially by using a representation similar to the gauge
P representation formally introduced here. The basic differences were the following.

\begin{enumerate}
\item Instead of a complex gauge \( \Omega  \), a phase factor \( e^{i\theta } \)
with a real \( \theta  \) variable, was used. 
\item The normalization with respect to the behavior of \( \theta  \) was
carried out explicitly inside the kernel, rather than post-simulation
in the moments as in Eq.~(\ref{moments}). 
\end{enumerate}
This type of representation is a norm-preserving gauge P representation,
as discussed earlier. A parametrized family of gauges led to stable
trajectories (as opposed to the large sampling error present with
a positive-P simulation). However, some systematic errors were seen
due to boundary terms. These boundary terms occurred because of the
stochastic growth of the gauge term in \(\Omega\) space, when
\( \theta  \) approached \( \pm \pi /2 \). With the gauge P representation
introduced in this paper, a wide range of gauges do not lead to any
systematic errors \cite{gauge_paper}, provided gauge growth is controlled.

We note here that the norm-preserving gauges have the property that,
in the present notation, \( g_{k}=i[1-i\Omega '']f_{k} \) . However,
while the growth of \( \Omega ' \) is stabilized, there is growth
in the variance of \( \Omega '' \). This means that the function
\( f_{k} \) must behave as a decreasing function of \( \Omega '' \)
in order to ensure that the distribution is bounded sufficiently in
the weight-function space to avoid finite boundary terms. The detailed requirements and conditions
for this type of gauge will be treated elsewhere.

\end{document}